\documentstyle[11pt,aaspp4]{article}
\begin{document}
\newcommand{\etal}{et~al.\ }
\newcommand{\lya}{Ly$\alpha$}
\newcommand{\cmss}{{\rm cm\: s^{-2}}}
\newcommand{\ergcmcms}{{\rm erg\:cm^{-2}\:s^{-1}}}
\newcommand{\ergs}{{\rm erg\:s^{-1}}}
\newcommand{\kms}{{\rm km\:s^{-1}}}
\newcommand{\hpc}{h_{65}^{-1}\:{\rm pc}}
\title{
Stellar Atmospheres Near an AGN:
\\
The Importance of Radiation Pressure from Trapped Lyman-$\alpha$ Photons
}
\author{Weihsueh A. Chiu}
\affil{Joseph Henry Laboratories, Department of Physics, Princeton, NJ 08544}
\and
\author{B. T. Draine}
\affil{Princeton University Observatory, Princeton, NJ 08544}

\author{Submitted to {\it The Astrophysical Journal}}

\begin{abstract}
  We derive an analytic expression for the intensity of 
resonance-line radiation ``trapped'' in a semi-infinite medium.
Given a source function and destruction probability
per scattering, 
the radiation pressure due to  
trapped photons can be calculated by
numerically integrating over analytic functions.
We apply this formalism to a plane-parallel model stellar atmosphere to 
calculate the radiation pressure due to Lyman-$\alpha$ photons 
produced following absorption of UV and X-rays from an AGN.
For low surface gravity stars near the AGN 
($g \sim 10\:\cmss,\ r \sim 0.25 {\rm\:pc}$), 
we find that the pressure due to 
Lyman-$\alpha$ photons becomes an appreciable fraction of that
required for hydrostatic support.  
If the broad emission line emitting gas in AGNs and QSOs 
consists of stellar outflows, it may be driven, in part, 
by Lyman-$\alpha$ pressure.

\end{abstract}


\section{Introduction}
\label{intro}

	Interest in the effect of hard UV and X-ray radiation on 
stellar atmospheres and stellar evolution began in earnest
with work on X-ray binaries (e.g., \cite{DOst73}, 
\cite{BSun73}, and \cite{Arons73}), but the subject 
has since been revisited in
the context of active galactic nuclei (AGN).  

	Consider a star of effective temperature $T_{\rm eff}$
and an AGN of luminosity $L = 10^{46}L_{46}\:\ergs$.
The incident power per area from the AGN is equal to
$\sigma T_{\rm eff}^4$ at the ``heating'' distance
\begin{equation}
d_h = \left(\frac{L}{4\pi\sigma T_{\rm eff}^4}\right)^{\onehalf} = 
	1.5\times 10^{17} L_{46}^{\onehalf}\left(
		\frac{5000 {\rm K}}{T_{\rm eff}}\right)^2
	\:{\rm cm};
	\label{eq:rheat}
\end{equation}
heating of the stellar surface by the AGN is important at distances
$d \lesssim d_h$.  Fabian (1979) was apparently the first to suggest 
that heating by AGN radiation might substantially affect
the envelopes of nearby stars, resulting in enhanced mass loss.
Edwards (1980) argued that when a star approached within
$d \lesssim d_h$, the irradiated photosphere would develop supersonic
horizontal winds carrying heat from the illuminated hemisphere
to the shadowed side.  Edwards conjectured that enhanced mass loss
could then occur, resulting in a ``cometary star'' as the stellar
outflow is accelerated radially away from the AGN\@.  
Matthews (1983) noted the importance of direct radiation pressure
from the AGN, arguing that this could ablate matter tangentially from 
the stellar photospheres of giant or supergiant stars.

	Voit \& Shull (1988) discussed the effects of X-rays
on the upper atmosphere of a star near the AGN\@.  They argued 
that X-ray heating of a $\sim 1 M_{\sun},$ $R_{*} 
\approx 100 R_{\sun}$ star ($g\approx 3\:\cmss$) would result in a 
hot, ionized wind, with
temperature at the critical point $T \approx 3\times 10^5 {\rm K}$;
radiation pressure in C IV, N V, and O VI resonance lines would 
contribute to the acceleration of the wind.  They concluded that stars 
which are already red supergiants could develop winds with
$\dot M \approx 10^{-7}L_{46}^{0.9}d_{18}^{-2}R_{*,100}^2 M_{\sun}
{\rm yr^{-1}}$, where the distance from the AGN is 
$d = 10^{18} d_{18}{\:\rm cm}$, and $R_{*,100} \equiv R_*/100R_{\sun}$.  
These mass loss rates exceed normal mass loss 
rates for red supergiants only within a distance of 
$d_{18} \lesssim 0.3 L_{46}^{0.45} R_{*,100}\:{\rm cm}$.  
They estimated the mass loss due to
ablation by radiation pressure (\cite{Matthews83}) for a
$\sim 1 M_\odot$ star to be 
$\dot M \approx 10^{-8}L_{46}d_{18}^{-2}R_{*,100}^{5/2}$
provided $d_{18} < 0.6L_{46}^{\onehalf} R_{*,100}$.

	There has also been some exploration of how AGN radiation
might affect the evolution of stars (\cite{Matthews83};
\cite{Verbuntetal84}; \cite{Tout89}).  Tout \etal discussed
the evolution of stars immersed in a blackbody background with 
temperature $T = 10^{3.75 - 4} {\rm K}$, concluding that
the main sequence evolution is largely unaffected, but that the
star will expand to a radius $\sim 10$ times larger than usual
during the ``red giant'' phase of evolution.
Note, however, that the energy density of AGN radiation equals
a $10^4{\rm K}$ blackbody at a distance of only $2\times 10^{16}
L_{46}^{0.5}\:{\rm cm}$ from the AGN\@.  Furthermore, it 
is not clear how their conclusions would have to be modified
for the anisotropic and nonthermal radiation field of the AGN\@.

	These discussions of the effects of AGN radiation 
on stellar atmospheres and evolution have led some to 
consider a ``bloated stars scenario'' (BSS) 
to explain the origin of the AGN broad line region (BLR). 
The BSS proposes that the BLR consists of the stellar winds and/or 
expanded envelopes of stars irradiated with AGN radiation
(e.g., \cite{Pen88}, \cite{Kaz89}).  The attractiveness of this
scenario stems from its efficient use of known resources: 
stars are believed to be present near the ``central engines''
of AGN, their gravity provides a possible 
``containment'' mechanism for BLR clouds, there is ample 
mass with which to replenish the clouds, and the mass shed 
provides material to fuel the AGN\@.  So far, however, 
no specific model has successfully explained how all these 
mechanisms might function, though pieces of the puzzle have
been explored (\cite{AN94}).  High signal/noise measurements
of line profiles indicate that the number of discrete clouds
or bloated stars must be large, perhaps exceeding $\sim
3\times 10^6$ for Mrk~335 (Arav \etal 1997) and $\sim
3\times 10^7$ for NGC~4151 (Arav \etal 1998).  Such
large numbers appear to challenge the BSS, 
but the difficulties may not be insurmountable.

	Recent observations and photoionization models of 
Baldwin \etal (1996) and Ferland \etal (1996) 
offer some concrete evidence
for the BSS\@.  They find that the BLR seems to contain 
several distinct components.  One (component ``A'') has sharp 
(FWHM $\approx 1000\:\kms$), symmetric line 
profiles centered on zero velocity, while another (component ``B'') 
appears to be outward-flowing, peaked at zero velocity but with a long
blue tail (down to $-11,000\:\kms$).  
They interpret ``A'' as the expanded envelopes 
of stars, and ``B'' as their radiatively accelerated, outflowing winds.

	The aim of the present paper is to call 
attention to the fact that irradiation of a
star by hard X-rays from the AGN will result in the generation of
Lyman-$\alpha$ (\lya) photons within the stellar atmosphere,
and the pressure of these trapped resonance-line photons may
contribute to mass loss.  The possible importance
of \lya\ photons was suggested
previously by Puetter [unpublished, cited in Penston (1988)
and Voit \& Shull (1988)].  Voit \& Shull rejected Puetter's suggestion,
arguing that the \lya\ pressure could be estimated
to be
\begin{equation}
	P_{\rm Ly\alpha} \approx 
	\left(\frac{aT^4}{3}\right)\frac{W_\lambda}{\lambda}
\label{eq:PVoitShull}
\end{equation}
where $W_\lambda$ is the equivalent width of the \lya\ absorption
profile of the gas between the point of interest and the surface,
and $T$ is the gas temperature.  Voit \& Shull then used
$W_\lambda/\lambda < 1$ to obtain an upper limit $P_{\rm Ly\alpha} 
\lesssim aT^4/3$
which, for stars of interest, is insufficient to ``bloat'' the atmosphere.
However equation (\ref{eq:PVoitShull}) presumes the radiation field within
the gas to be close to a blackbody at the gas 
temperature $T$ --- an assumption
which need not be valid when \lya\ photons are being generated
within the gas by nonthermal processes, such
as H(1s$\rightarrow$2p) excitation by photoelectrons and secondary
electrons in a non-Maxwellian ``tail'' to the electron energy 
distribution function.  It is therefore essential to
estimate the rate of production of \lya\ photons, and to
examine their diffusion in physical space and frequency space
as well as the possibility of photon ``destruction'' by,
for instance, collisional de-excitation of electronically excited
H atoms.  

	In \S\ref{sec-linetrans}, we give a brief treatment of 
resonance-line transfer, 
extending the detailed calculations of Neufeld (1990).  In
particular, we derive an expression for the case of a semi-infinite, 
absorbing slab of material.  Since resonance-line photons created in a stellar
atmosphere will almost certainly not be able to scatter through the
entire star without being absorbed, this limit is an excellent 
approximation.  In \S\ref{sec-model}, we apply our calculational  
method to a simple,
plane parallel, irradiated model atmosphere.  
In \S\ref{sec-results}, 
we present the results. In \S\ref{sec-summ} 
we summarize and discuss implications and further directions for work.

\section{Resonance Radiation Transfer}
\label{sec-linetrans}

	Most current photoionization treatments, including
CLOUDY (Ferland 1996), use the
escape probability approximation to address resonance-line
radiation transfer.  For example, Elitzur and Ferland (1986)
use the escape probability approximation (EPA) to 
derive the frequency-integrated radiation intensity ${\cal J}_{ul,\:\rm EPA}$
for a resonance line at frequency $\nu_{ul}$, corresponding to a 
transition from level $u$ to level $l$, to be
\begin{equation}
{\cal J}_{ul,\:\rm EPA} = B(\nu_{ul},\:T_{\rm exc})\Delta\nu_{ul},
\end{equation}
where $B(\nu_{ul},T_{\rm exc})$ is the Planck function at the 
line excitation temperature (which is calculated from the 
level populations) and $\Delta\nu_{ul}$ is the line width. 
The radiation pressure $P_{ul,\:\rm EPA}$ is given, as usual, by 
\begin{equation}
P_{ul,\:\rm EPA} = \frac{4\pi}{3c} {\cal J}_{ul,\:\rm EPA}.
\label{eq:pradrealdef}
\end{equation}  
In this purely local treatment, 
the escape probability approximation assumes that a resonance photon
at a point $r$ has a probability $p_{ul}$ per scattering of ``escaping'' 
to a defined boundary $r_0$, where it is assumed photons free-stream
to infinity.  
Thus, the differential radiation pressure $P^{\rm diff}_{ul,\:\rm EPA}(r)$
at any point $r$ is given by
\begin{eqnarray}
P^{\rm diff}_{ul,\:\rm EPA}(r) & \equiv & P_{ul,\:\rm EPA}(r) -
	P_{ul,\:\rm EPA}(r_0) 
\label{eq:p_escprob1}\\
& = & P_{ul,\:\rm EPA}(r) \times (1-p_{ul})\\
& = & \frac{4\pi}{3c}B(\nu_{ul},\:T_{\rm exc})\Delta\nu_{ul} (1-p_{ul}).
\label{eq:p_escprob2}
\end{eqnarray}

Confusingly, it is this {\it differential} radiation pressure which
Elitzur and Ferland (1986) and CLOUDY (Ferland 1996) 
refer to as the ``line radiation pressure.''  We
will always use the term ``pressure'' and the symbol $P_i$ 
for the physical pressure (momentum flux) at point $r$ 
due to particles $i,$
such as gas particles or resonance line photons.  We will use the
term ``differential pressure'' and the symbol 
$P_i^{\rm diff} \equiv P_i(r)-P_i(r_0)$
for the pressure difference due to particles $i$ 
between the point of interest $r$ 
and the boundary point $r_0$.  

Elitzur and Ferland (1986) note that the line width 
$\Delta\nu_{ul}$ is the ``crucial
parameter'' in this calculation of the radiation pressure.  
In particular, it is 
not well understood how to calculate $\Delta\nu_{ul}$ at extremely
large optical depths when the escape probability is comparable 
to the ``destruction probability'' per scattering,
for example due to collisional de-excitation.

	However, the case of very large optical depth
is precisely that which Neufeld (1990) considers 
using a completely different
approach.  He presents in \S\S 2--3 of his paper 
an {\it exact} and {\it analytic} 
calculation derived from Fourier transforming
a diffusion equation.  He solves this equation 
for a unit delta-function photon source in a
uniform, optically thick (but finite) slab,
with and without a finite destruction probability.  
In particular, at optical depth $\tau$ with a source at $\tau_s$, 
the angle-averaged intensity $J(\tau,\tau_s,x)$ and 
frequency-integrated intensity ${\cal J}(\tau,\tau_s)$
are given in terms of infinite series [e.g., his equation (2.7)].
Unfortunately, these series converge rather slowly.
However, we show below that if we take the limit of a semi-infinite
slab, then the angle-averaged intensity $J(\tau,\tau_s,x)$ can
be expressed in terms of analytic functions which can be 
integrated numerically with relative ease. 
Below, we first present the notational conventions we use, and then
proceed with our derivation of the intensities $J$ and ${\cal J}$ in
the case of a semi-infinite slab.

Neufeld (1990) defines $a$ as the ratio of the natural line width to 
the Doppler width and $\tau$ as the mean optical depth:\footnote{
For consistency, we retain the definitions of $\tau$ and 
$\sigma_\alpha$ used by Neufeld (1990), Harrington (1973),
and other workers: $\sigma_\alpha = \sqrt{\pi} \sigma_\circ,$
where $\sigma_\circ$\ is the absorption cross section
at line-center.}
\begin{equation}
	a = \frac{\gamma}{4\pi}\frac{\lambda_0}{\sqrt{2}\,v_{1d}}, \quad
	\tau = 	\sigma_{\alpha} N_0, \quad
	\sigma_\alpha \equiv  
		\frac{\pi e^2 f_{lu}\lambda_0}{m_e c\sqrt{2}\,v_{1d}}
	\label{eq:a_tau_def} 
\end{equation}
where $\gamma$ is the decay rate, $\lambda_0$ is the
line-center wavelength,
$v_{1d} = \sqrt{kT/m_{\rm H}}$ 
is the one-dimensional velocity dispersion,
$m_{\rm H}$ is the mass of atomic hydrogen,
$f_{lu}$ is the oscillator strength,  
$m_e$ is the electron mass, 
$N_0$ is the column density of absorbers,
and the other symbols have their usual meanings.  
For \lya, this corresponds to 
$v_{1d} = 9.082\:T_4^{\onehalf}\:\kms$,
$a = 4.717\times 10^{-4} T_{4}^{-\onehalf}$,
and $\tau = 1.045\times 10^{-13} T_{4}^{-\onehalf} N_0\:{\rm cm}^2$, where
$T_4$ is the temperature in units of $10^4\: {\rm K}$.
The dimensionless frequency variables $x$ and $\sigma(x)$ which
Neufeld uses are given by:
\begin{equation}
	x \equiv \frac{\lambda_0(\nu-\nu_0)}{\sqrt{2}\,v_{1d}}, \quad
	\sigma(x) \equiv \sqrt{\frac{2}{3}}\int_0^x \frac{dx'}{\phi(x')},
	\label{eq:xsigdef}
\end{equation}
where $\phi(x)$ is the Voigt profile normalized such that
$\int\phi(x)dx = 1$.  In this convention, $\phi(x)$ takes
on the limits
\begin{equation}
	\phi(0)= \frac{1}{\sqrt{\pi}}\left(1 + O(a)\right);\quad 
	\phi(x) \approx \frac{a}{\pi x^2}, x \gg 1.
	\label{eq:phiapprox}
\end{equation}
The intensity $J(\tau,\tau_s,x)$ is 
given per unit frequency in $x$, so that the frequency-integrated
intensity is given by
\begin{equation}
{\cal J}(\tau, \tau_s) \equiv \int J(\tau,\tau_s,x) dx.
	\label{eq:jintdef}
\end{equation}

	Since Neufeld's solutions for the intensities $J(\tau,\tau_s,x)$
and ${\cal J}(\tau,\tau_s)$ are derived for a delta-function source, 
they act as Green's functions for resonant photon sources.  Consider
a slab of total thickness $\tau_0$.
For a \lya\ emissivity per unit volume $4\pi j$, 
we define the injected power per unit area per unit optical depth as
\begin{equation}
\xi(\tau_{s}) = \frac{4\pi j({\rm Ly}\alpha)}{n({\rm H^0})\sigma_\alpha}.
	\label{eq:xidef}
\end{equation}
The excitation temperature $T_{\rm exc}$ at optical depth $\tau$ is found by 
integrating $J$ near line center over all source locations
$\tau_{s}$, weighted by $\xi$, to
obtain the total intensity at line center, and comparing with a 
blackbody intensity:
\begin{equation}
	B(\nu_0,T_{\rm exc}) \frac{d\nu_0}{dx} = \int_0^{\tau_0} 
		\xi(\tau_s)J(\tau,\tau_s,0)d\tau_s,
	\label{eq:texcdef}
\end{equation}
where $B(\nu_0,T_{\rm exc})$ is the usual Planck function at the
excitation temperature $T_{\rm exc}$, evaluated at line center.  
Since the majority of
scatterings occur near line center, the local 
level population for \lya\ should be given by
\begin{equation}
	\frac{n_{\rm 2p}}{n_{\rm 1s}} = \frac{g_{\rm 2p}}{g_{\rm 1s}}
		\exp\left(-\frac{\chi_{21}}{kT_{\rm exc}}\right),
	\label{eq:levelpopdef}
\end{equation}
where the $n$'s are the level populations, $g$'s are the statistical
weights, and $\chi_{21}=10.2\, \mbox{eV}$.  
The radiation pressure 
$P_{\rm rad}(\tau) = 4\pi{\cal J}/(3c)$ 
is given similarly by 
\begin{equation}
P_{\rm rad}(\tau) =  \frac{4\pi}{3c}\int_{0}^{\tau_0}
	\xi(\tau_{s}){\cal J}(\tau,\tau_{s})d\tau_{s}.
	\label{eq:Praddef}
\end{equation}

	Consider photons created at line center with a
constant probability per scattering 
$\epsilon_0 \ll 1$
that the photon will be ``destroyed'' by, for instance, 
collisional de-excitation.  Using Neufeld's equations (2.7) and
(3.11)--(3.13), we find that for a slab of total thickness 
$\tau_0$ and source at $0 \le \tau_s \le \tau_0$, the 
angle-averaged intensity at $0 \le \tau \le \tau_0$ can
be written as
\begin{eqnarray}
J(\tau,\tau_s,x) & = & \frac{\sqrt{6}}{8\pi^2}
	\left\{F\left[\tau - \tau_s,\sigma(x)\right]
	- F\left[\tau + \tau_s + \frac{4}{3\phi(x)}
			+ O\left(\tau_0^{-1}\right),
			\sigma(x)\right]
	\right\},
	\label{eq:Jdef}
	\\
F[v,\sigma(x)] & \equiv & \frac{\pi}{\tau_0} 
	\sum_{n = 1}^{\infty}
	\frac{\cos(\mu_n v)}{\mu_n +
		\epsilon_0\sqrt{6}/2} \exp[-\mu_n |\sigma(x)|],
	\label{eq:Fdef}
	\\
\mu_n & = & \frac{n\pi}{\tau_0}
	\left[1 + O\left(\tau_0^{-1}\right)\right].
	\label{eq:mudef}
\end{eqnarray}
Note that our $\tau_0$ is {\it twice} Neufeld's $\tau_0$.
 
In a stellar atmosphere, we essentially have a semi-infinite slab, where
$\tau_0 \rightarrow \infty$.  
We can therefore neglect the terms of order $\tau_0^{-1}$, and 
replace the sum in equation (\ref{eq:Fdef}) with an integral.  
With the substitutions
\begin{equation}
\tau_d \equiv \frac{2}{\epsilon_0 \sqrt{6}},
	\label{eq:tauddef}
\end{equation}
\begin{equation}
u_n  \equiv  \tau_d \mu_n = n\pi\tau_d/\tau_0, \quad
\frac{du_n}{dn}  = \pi\tau_d/\tau_0,
	\label{eq:udef}
\end{equation}
$F$ then becomes
\begin{eqnarray}
F(v,\sigma)
	& \rightarrow & 
	\int_0^\infty
	\frac{\cos(u\,v/\tau_d)}{1 + u}
		\exp[-u|\sigma(x)|/\tau_d] du
	\label{eq:flim} \\
	& = & {\cal E}(z) \equiv {\rm Re}[e^z E_1(z)],
	\label{eq:flimdef}\\
	z & \equiv & \frac{|\sigma(x)| + i|v|}{\tau_d},
	\label{eq:zdef} 
\end{eqnarray}
where $E_1(z)\equiv \int_z^\infty w^{-1} e^{-w} dw$ 
is the exponential integral with complex argument.
The quantity $\tau_d$ is the effective optical depth due to finite
destruction probability.  

Therefore, the angle-averaged intensity in the limit that
$\tau_0\rightarrow \infty$ is given by
\begin{equation}
	J(\tau,\tau_s,x) = \frac{\sqrt{6}}{8\pi^2}
	\left\{{\cal E}\left[\frac{|\sigma(x)| + i|\tau
		-\tau_s|}{\tau_d}\right]
	- {\cal E}\left[\frac{|\sigma(x)| + i|\tau+\tau_s+\frac{4}{3\phi(x)}|}
		{\tau_d}\right]\right\}.
\label{eq:jgreenfunc}
\end{equation}
At line center, where $x=\sigma=0$ and the argument $z$ is purely
imaginary, ${\cal E}(z)$ can be expressed
in terms of cosine and sine integrals 
$\mbox{Ci}(y) \equiv -\int_y^\infty t^{-1}\cos t\, dt$ 
and $\mbox{si}(y) \equiv -\int_y^\infty t^{-1}\sin t\, dt$:
\begin{equation}
{\cal E}(i y) = 
	-\left[\cos y\ \mbox{Ci}(y) + 
		\sin y\ \mbox{si}(y)\right].
\label{eq:efuncimag}
\end{equation}
Using equations (\ref{eq:jgreenfunc})--(\ref{eq:efuncimag}) in
equation (\ref{eq:texcdef}) gives the excitation temperature, and
hence the level population at each point in the slab using 
equation (\ref{eq:levelpopdef}).

To find the frequency-integrated intensity ${\cal J}$, we 
must integrate $J$ over the frequency $x$, 
equation (\ref{eq:jintdef}).  
The exact expression for the frequency-integrated intensity 
in the limit that $\tau_0\rightarrow\infty$ is thus:
\begin{eqnarray}
{\cal J}(\tau,\tau_{s}) & = & 
	\frac{\sqrt{6}}{8\pi^2}
	\left[{\cal F}_{\tau_d}^{}\left(|\tau - \tau_s|\right) -
		{\cal G}_{\tau_d}\left(\tau + \tau_s\right)
	\right],
	\label{eq:Jlimit}
	\\
{\cal F}_{\tau_d}(v) & \equiv & 2\ \times\  
	\int_0^{\infty} 
		{\cal E}\left[\frac{\sigma(x) + iv}{\tau_d}\right] dx,
	\label{eq:Fcaldef}
	\\
{\cal G}_{\tau_d}(v) & \equiv & 2\ \times\ 
	\int_0^{\infty} 
		{\cal E}\left[\frac{\sigma(x) + 
			i\left(v+\frac{4}{3\phi(x)}\right)}{\tau_d}\right] dx
	\label{eq:Gcaldef}.
\end{eqnarray}
The functions ${\cal F}$ and ${\cal G}$ in general need to be
calculated numerically.  Obtaining the numerical integral is
simplified by taking into account the following limits 
of the integrand:
\begin{equation}
{\cal E}\left[\frac{\sigma(x)+iv}{\tau_d}\right] \propto
\cases{\sim\mbox{constant}&if $x \lesssim (av)^{\onethird}$ and 
				$x \lesssim (a\tau_d)^{\onethird}$;\cr
	\sim x^3&if $(a\tau_d)^{\onethird} \lesssim 
				x \lesssim (av)^{\onethird}$ 
				and $\tau_d < v$;\cr
	\sim x^{-3}&if $x \gtrsim (a\tau_d)^{\onethird}$ 
				and $x \gtrsim (av)^{\onethird}$.\cr}
\end{equation}

If the \lya\ optical depths $\tau$ and $\tau_s$, as
well as the effective optical depth to destruction $\tau_d$, 
are all large, so that 
$(av)^{\onethird} \gg 1$\ and $(a\tau_d)^{\onethird} \gg 1$, then
the scattering is dominated by the damping wings.  We may 
then approximate the Voigt profile $\phi(x)$ by using equation  
(\ref{eq:phiapprox}) and $\sigma(x)$ by 
$\sigma(x) \approx \sqrt{2/3}\,\pi x^3/(3a)$.  
This approximation leads to an {\it analytic} expression 
for ${\cal F}$ and ${\cal G}$: 
\begin{eqnarray}
{\cal F}_{\tau_d}(v) & \approx & \frac{2\sqrt{3}}{3}
	\Gamma\left(\onethird\right)\Gamma\left(\twothirds\right)
	\left(\frac{a\tau_d}{\sqrt{2}\, \pi}\right)^{\onethird}
	{\rm Re}\left[\exp(iv/\tau_d)
		\Gamma\left(\onethird,iv/\tau_d\right)\right]\\
{\cal G}_{\tau_d}(v) & \approx & {\cal F}_{\tau_d}
	\left[v + \frac{4}{3\phi(0)}\right],
\end{eqnarray}
where the incomplete gamma function with 
an imaginary argument is $\Gamma(b,iy) \equiv i^{b}\int_{y}^{\infty} 
t^{b-1}e^{-it}dt$.  This case is not applicable to the AGN-illuminated
stellar atmosphere model we discuss below, but we include it for
completeness.

\section{A Stellar Atmosphere Near an AGN}
\label{sec-model}

	Our procedure for evaluating the importance of 
\lya\ pressure in a stellar atmosphere near an AGN is as
follows:
\begin{enumerate}
\item[A.] Use a slightly-modified version of CLOUDY (\cite{Cloudy90}) to 
calculate the thermal and 
ionization structure of a plane-parallel, gravitationally
confined gas illuminated by an AGN\@.  
\item[B.] Calculate the \lya\ production and destruction
rates from A.
\item[C.] Use B and the analytic model from \S 2 to calculate 
the \lya\ radiation pressure $P_{{\rm Ly}\alpha}(r)$.
\item[D.] Compare the differential radiation pressure 
$P_{{\rm Ly}\alpha}^{\rm diff} \equiv 
P_{{\rm Ly}\alpha}(r) - P_{{\rm Ly}\alpha}(r_0)$ 
from C with $P_{\rm hyd}$, the pressure required for hydrostatic 
equilibrium.
\end{enumerate}
If $P_{{\rm Ly}\alpha}^{\rm diff}\approx 
\frac{1}{2}P_{\rm hyd}$, then the atmosphere
may be unstable, especially since
other resonance lines will likely contribute to the total radiation
pressure (Elitzur \& Ferland 1986).  
We describe each of these steps in greater detail below.

\subsection{Use of CLOUDY} 
\label{sec-physics}

	We have modified the CLOUDY program
(version 90.02) of Ferland (1996) to model 
a plane-parallel atmosphere with a constant gravitational acceleration, 
aligned normal to the incident AGN flux.  
We ignore the radiation from the star, considering only
the effect of the AGN illumination on a gravitationally
confined gas.  The modifications to CLOUDY thus only involve adding 
a constant gravitational acceleration $g$ to the equation for the total 
pressure equilibrium:
\begin{equation}
P_{\rm tot}(r) \equiv P_{\rm gas}(r) + P_{\rm lines}(r)
	= \int_{r}^{r_0} [a_{\rm cont}(r) + g]\rho(r) dr
	+ P_{\rm gas}(r_0) + P_{\rm lines}(r_0).
\label{eq:pequil}
\end{equation}
Here $r$ is the distance from the center of the star, $r_0$ is 
the starting point of the integration ($r \leq r_0$), 
$P_{\rm gas}(r)$ and $P_{\rm lines}(r)$ are 
the pressures due to gas and resonance lines 
respectively (the latter calculated by CLOUDY 
using escape probabilities), and $\rho(r)$ is the gas mass density.
We define the hydrostatic pressure $P_{\rm hyd}$ to be 
the total pressure $P_{\rm tot}$ minus the free-streaming radiation 
pressure at $r_0$, $P_{\rm lines}(r_0)$:
\begin{equation}
P_{\rm hyd}(r)  \equiv  P_{\rm tot}(r) - P_{\rm lines}(r_0).
\label{eq:p_hyddef}
\end{equation}
This hydrostatic pressure provides the basis 
of comparison with the differential
radiation pressure $P_{\rm lines}^{\rm diff}$, since it is the
equivalent gas pressure necessary for hydrostatic equilibrium,
in the absence of any line radiation pressure.  
The acceleration due to the momentum deposition from the 
attenuation of the AGN continuum flux is 
given by 
\begin{equation}
a_{\rm cont}(r) = \frac{1}{c \mu_{\rm H} m_{\rm H}}\int_{-\infty}^\infty 
	\sigma_\nu^{\rm eff}(r) f_\nu(r) d\nu,
\end{equation}
where $\mu_{\rm H}$ is the gas mass per H nucleon in units of the
mass of atomic hydrogen $m_{\rm H}$, 
$f_\nu(r)$ is the AGN energy flux per unit frequency,
and $\sigma_\nu^{\rm eff}(r)$ 
is the effective cross section per H nucleon, the latter two
quantities evaluated at frequency $\nu$ and distance $r$.
Recall that the line radiation pressure calculated by CLOUDY 
$P_{\rm Ly\alpha}^{\rm CLOUDY}(r)$ is actually the
differential line radiation pressure $P_{\rm Ly\alpha}^{\rm diff}(r)$.
Our treatment, on the other hand, explicitly calculates 
both $P_{\rm Ly\alpha}(r)$ and $P_{\rm Ly\alpha}(r_0)$, the 
momentum flux in free-streaming resonance photons at $r_0$, 
and takes their difference to find $P_{\rm Ly\alpha}^{\rm diff}(r)$.

The abundances we use are those 
used by \cite{F96} and \cite{B96}: 
$\log_{10}$(H: He: Li: Be: B: C: N: O: F: Ne: Na: Mg: Al: Si: P: 
S: Cl: Ar: K: Ca: Sc: Ti: V: Cr: Mn: Fe: Co: Ni: Cu: Zn) = 
(0: --0.907: --7.92: --9.82: --8.35: --3.08: --2.94: --2.23: --6.71: --3.10: 
--4.87: --3.6: --4.71: --3.64: --5.62: --3.97: --5.92: --4.63: --6.06: 
--4.85: --8.76: --6.91: --7.83: --6.16: --6.31: --4.18: --8.5: --5.6:
--7.58: --7.19).
The incident AGN flux we use is the same one used by 
\cite{B96} and \cite{F96} in their photoionization models of the
BLR\@.  Their formula is equivalent to
\begin{eqnarray}
\nu f_{\nu}^0 & = & \frac{\nu L_\nu}{4\pi d^2} =
	\frac{A}{4\pi d^2}\left(
		\frac{\nu}{\nu_0}e^{-\nu/\nu_c} + 
		0.09\:e^{-\nu_0/\nu}\right)
	\label{eq:fluxdist}\\
h\nu & \le & E_{\rm max} \nonumber
\end{eqnarray}
where $h\nu_{c} = 21.5\:{\rm eV}$ and  
$h\nu_0 = 13.6\:{\rm eV}$, which corresponds to a UV--X-ray spectral
index $\alpha_{ox} = -1.2$.  For a Hubble's constant 
$H_0 = 65\: h_{65}\:\kms\:{\rm Mpc^{-1}}$, 
Baldwin \etal (1996) adopt 
$A = 1.77\times 10^{47}\: h_{65}^{-2}\:\ergs$ 
and a distance of $d = 0.25\:\hpc$, both of which
we adopt as well.  
We use a high energy cutoff 
of $E_{\rm max} = 100\:{\rm keV}$.
The luminosity in ionizing radiation is related to 
equation (\ref{eq:fluxdist}) by 
\begin{equation}
L_{\rm ion} \equiv \int_{\nu_0}^{\nu_{\rm max}} L_\nu d\nu
	= 1.57 \cdot A.
\label{eq:incidentflux}
\end{equation}
Using $L_{\rm ion}$ as an estimate for the AGN 
luminosity heating of
the star, we obtain using equation (\ref{eq:rheat}) 
a ``heating distance'' of 
\begin{equation}
d_{h} \approx \left(\frac{L_{\rm ion}}
	{4\pi\sigma T_{\rm eff}^4}\right)^{\onehalf} = 0.26 
	\left(\frac{5000\:{\rm K}}{T_{\rm eff}}\right)^2 \hpc.
\end{equation}
Therefore, for a $T_{\rm eff}=5000\:{\rm K}$ star, our calculations
are at the heating distance.

Finally, a few notes about our integration procedure in CLOUDY\@. 
We begin our integration at a large distance 
with a total hydrogen density of 
$10^{10}\: {\rm cm^{-3}}$ [which also determines $P_{\rm gas}(r_0)$ in
equation (\ref{eq:pequil})] and an unattenuated incident flux
(\ref{eq:fluxdist}); we integrate inward and stop when the 
electron fraction
drops to 5\%; and we iterate the solution until pressure
equilibrium (with CLOUDY's implementation of the
resonance-line radiation pressure) is reached.

\subsection{Calculation of Lyman-$\alpha$\ Production
and Destruction Rates}
\label{sec-prod_dest_calc}

	Given the ionization equilibrium calculated by CLOUDY, we
calculate the \lya\ production and destruction rates in the 
following manner.  For \lya\ production, we assume all
electrons entering either $n=2$ states which do not involve the
absorption of a scattered \lya\ photon result in the 
production of a {\it new} \lya\ photon.  We are justified in 
including electrons entering the 2s state, because
at the densities we consider, an electron in the 2s state has
an overwhelming probability of being collisionally transferred
to the 2p state, and because once an electron is 
in the 2p state, the overwhelming probability is that
it will result in the emission of a \lya\ photon.
Thus the production processes we consider  
include (1) radiative and collisional (3-body) recombination
directly to $n=2,$ (2)
photoexcitation by the AGN continuum, (3) collisional excitation
by thermal electrons,
(4) radiative and collisional
de-excitation from $n>2$, and (5) collisional excitation by
photoelectrons and secondary electrons.
This gives a \lya\ production 
emissivity per unit volume, as a function of hydrogen
column density $N_{\rm H}$, of 
\begin{equation}
4\pi j(N_{\rm H})   =  \chi_{21} n_{\rm H} \sum_{j=1}^{5} 
	\Gamma_j({\rm \rightarrow 2p,2s})
	\label{eq:Lyemiss}
\end{equation}
where $\chi_{21}$ is $10.2\: {\rm eV}$, $n_{\rm H}$ is the total hydrogen 
number density, $\Gamma_j$ is the rate for process $j$ 
per hydrogen nucleon.  In Figure 1, we show the 
\lya\ photon production rate per hydrogen nucleon 
[i.e., the summation in equation (\ref{eq:Lyemiss})]
in the case of stellar atmospheres with 
$g=1,\ 10,\ {\rm and}\ 100\:\cmss$.

	The destruction probability $\epsilon_0$ is the 
probability per scattering that either an electron in 2p does 
not end up producing a (scattered) \lya\ photon, or that 
a scattered \lya\ photon is absorbed by something other
than an electron in the 1s state of hydrogen.  
Thus the processes we consider 
are (1) photo- and collisional ionization from 2p, (2) collisional
de-excitation from 2p to 1s, (3) photo- and collisional excitation 
from 2p to $n>2$, (4) absorption by 
the ${\rm H}_2$ Lyman bands from $v=2$, $J=5,6$, 
and (5) transition to 2s followed by (5a) 
two-photon decay to 1s, (5b)  photo- and collisional ionization 
from 2s, (5c) collisional de-excitation from 2s to 1s, (5d) 
photo- and collisional excitation from 2s to $n>2$.
For molecular hydrogen, we use the rates and population fractions
from Neufeld (1980, section V, {\it a, ii--iii}). 
We therefore obtain a destruction probability per scattering of:
\begin{eqnarray}
\nonumber
\epsilon_0 & = & 1.2\times 10^{-3}\frac{n_{\rm H_2}}{n_{\rm H}}+
 	A_{21}^{-1} \times \bigg[
	\Gamma_1({\rm 2p\rightarrow})
	+ \Gamma_2({\rm 2p\rightarrow})
	+ \Gamma_3({\rm 2p\rightarrow}) + \\
	& &\Gamma(\rm 2p\rightarrow 2s) 
		\frac{\Gamma_{5a}(\rm 2s\rightarrow) 
		+ \Gamma_{5b}(\rm 2s\rightarrow) 
		+ \Gamma_{5c}(\rm 2s\rightarrow) 
		+ \Gamma_{5d}(\rm 2s\rightarrow)}
		{\Gamma(\rm 2s\rightarrow 2p)} 
	\bigg]
\label{eq:epsdef}
\end{eqnarray}
where the first term is due to molecular hydrogen, 
$A_{21}$ is the \lya\ decay rate, $\Gamma_i({\rm 2p\rightarrow})$
is the rate per 2p atom for process $i$, 
$\Gamma_i({\rm 2s\rightarrow})$ is the rate per 2s atom of
process $i$, and 
$\Gamma({\rm 2p\rightarrow 2s})$ and $\Gamma({\rm 2s\rightarrow 2p})$ 
are the rates per 2p or 2s atom
of transition to 2s or 2p, respectively.  In Figure 2, we
show the \lya\ destruction probability per scattering which
we calculate using equation (\ref{eq:epsdef}) and the 
results of CLOUDY for $g=1,\ 10,\ {\rm and}\ 100\:\cmss$.
We note that we find \lya\ destruction to be dominated 
by photo- and collisional ionization and 
excitation out of the $n=2$ states.  

\section{Results of Calculation of Lyman-$\alpha$\ Pressure}
\label{sec-results}

	With the distribution of \lya\ sources 
from equations (\ref{eq:xidef}) and (\ref{eq:Lyemiss}), 
we numerically integrate equation~(\ref{eq:Praddef}) with
equations (\ref{eq:Jlimit})--(\ref{eq:Gcaldef}) and 
(\ref{eq:epsdef}) 
to determine $P_{\rm Ly\alpha}$ in each layer.  

	For the incident flux $A/(4\pi d^2) = 1.77\times 10^{47} 
\:h_{65}^{-2}\:\ergs/[4\pi (0.25\:\hpc)^2]$ 
estimated by \cite{B96} and \cite{F96} (corresponding 
to a luminosity $L_{46}=27.8\:h_{65}^{-2}$),
the only parameter that can be varied is the 
gravitational acceleration.  We consider $g=1,\ 10,\ 100\:\cmss$.
The results are presented in Figures 3--5.  It appears
that for $g\lesssim 10\:\cmss$, the \lya\ pressure rises to 
$\sim 10\%$ of the hydrostatic pressure.
In this region, the total hydrogen column density 
$N_{\rm H} \sim 10^{22}\:{\rm cm}^{-2}$.  
The gas number density is $n\sim 10^{11}\:{\rm cm}^{-3}$,
below the estimate by Ferland \etal (1996)
and Baldwin \etal (1996) of $n\sim 10^{12.5}\:{\rm cm}^{-3}$.  
We also show results for stellar atmospheres at 
$d = 0.125$\ and 0.5$\:\hpc$, corresponding to half and
twice the heating distance, and obtain similar results.
Our results are insensitive to 
changing the ``boundary'' hydrogen number
density by a factor of $\sim 10$.   
In Figure 6, we compare the $P_{{\rm Ly}\alpha}^{\rm diff}$ 
that we derive with that which CLOUDY calculates; note that 
the escape probability approximation used by CLOUDY 
tends to overestimate the radiation pressure, at least
in the cases examined here.  

	As a check, we compare the excitation temperature 
derived from CLOUDY's escape probability approximation with 
that derived from our own calculation using equation 
(\ref{eq:texcdef}).  The results, shown in Figure 7,
are quite consistent, implying that 
the equilibrium level populations are 
well described by the escape probability approximation.  
Hence, it is the scattering in the damping wings of the line profile, 
which does not significantly affect the level population but which
greatly affects the line width, which is inaccurately 
calculated by CLOUDY in the escape probability approximation.
In comparing the excitation temperature to the kinetic
temperature, we see from Figure~7 that there is a region
of column density $\sim 10^{22}\:{\rm cm}^{-2}$ in which
$T_{\rm exc}$ exceeds $T_{\rm kin}$ by a factor $\gtrsim 2$.
Hence, the naive estimate of the radiation pressure
using equation (\ref{eq:PVoitShull}) may be as much 
as a factor of $\sim 16$ too small.  

	Our objective has been to carry out an exploratory
investigation of the importance of \lya\ pressure in the
atmosphere of a star which is being irradiated by an AGN.
This study has involved a number of simplifying approximations.
In particular, we have treated the atmosphere as plane-parallel 
with normally-incident irradiation; this, of course, precludes
examination of the important streaming effects which appear
likely to ensue from the transverse pressure gradients which
must arise from the non-uniform irradiation of the stellar surface.

	In view of the neglect of the transverse pressure gradients
and resulting motions, we have not iterated to a fully self-consistent
solution.  Our hydrostatic equilibrium is obtained using CLOUDY's 
estimate for the radiation pressure rather than our more accurate
result for $P_{\rm Ly\alpha}$.  Nor have we addressed the issue of 
stability or included radiation pressure from lines other than
\lya\ and from the star's own luminosity.  

	With these numerous simplifications, 
we find that in low gravity stars, the differential pressure
due to \lya\ can rise to $\sim 10$\% of the hydrostatic pressure.	

\section{Summary and Discussion}
\label{sec-summ}

	We have developed a formalism for computing 
the \lya\ pressure in a plane-parallel stellar atmosphere.  
The treatment of resonance-line trapping given in \S 2 can be
incorporated into already developed stellar atmosphere codes.
For conditions believed representative of the BLR (ionizing flux 
of $\sim 4\times 10^{10}\:\ergcmcms$),
stars with $g\lesssim 10\:\cmss$ will 
develop an appreciable pressure due to trapped \lya\ 
photons in their atmospheres.

	We have neglected numerous complicating factors, all of which
appear likely to enhance the rate of mass loss.
First of all, the gravitational acceleration $g\propto r^{-2}$,
rather than the assumed $g=\mbox{constant}$.  As a result, the
outer layers of the atmosphere will have a larger scale height
and must ultimately go over into a thermal wind, even if the
lower layers are essentially hydrostatic and stable.  With the 
outer layers of the atmosphere photoionized and heated to
$T_4 \gtrsim 1$, a fluid element will have positive enthalpy 
at $r \gtrsim 600\, T_4^{-1} (M/M_{\sun}) R_{\sun}$.

	Furthermore, we have neglected the radiation pressure due
to resonance lines other than \lya, as well as the radiation 
from the star itself.  Ferland (1997) reports
that including radiation pressure from all resonance lines
rather than just a select few sometimes
increases the total radiation by an order of magnitude (using
CLOUDY's escape probability approximation, of course).  
Since no part of our derivation in \S 2 depends explicitly on 
the resonance line being \lya, it is straight forward to
extend our calculation to include other resonance lines.  

	Most important, however, is the fact that except along
the ``axis'' --- the line from the center of the star to the
AGN --- the atmosphere will be subject to substantial tangential
stresses due to three effects related to anisotropic irradiation
by the AGN:
\begin{enumerate}
\item The momentum deposited by absorbed photons --- at anywhere
other than the axis, the momentum will have a tangential
component;
\item Transverse gradients in the \lya\ pressure --- at any
particular radial column density $N_{\rm H}$, the \lya\ production
rate will be greater along the axis, where X-rays are entering
radially; and
\item Transverse gradients in the gas temperature --- the rate
of X-ray heating will be largest along the axis, where the
zenith angle is zero.  
\end{enumerate}
These tangential stresses must result in significant tangential
flows, with flow velocities that might approach escape speeds
along the ``terminator,'' perhaps resulting in a ``cometary star'' as
envisioned by Edwards (1980) and Matthews (1983).

	It would obviously be of great interest to calculate 
axisymmetric fluid-dynamical models of red giant atmospheres
subject to intense X-ray irradiation in order to investigate the
effects of tangential stresses. 
It should be noted, of course, that we
lack a quantitative theory for mass loss even from unperturbed
evolved stars, so we are hardly poised to develop a definitive
treatment including this new complication.  Nevertheless, it
does appear likely that trapped \lya\ may help drive mass loss 
in low-$g$ stars subject to intense X-ray irradiation.
The extra \lya\ radiation pressure and associated tangential
stresses might induce such evolving stars to ``bloat'' further and 
lose mass prematurely, so that  
the population of stars near the AGN core could be skewed from
the average throughout the galaxy.  

	Finally, we note that none of our calculations are
dependent on the gravitating object being a star.  Recently, 
Walker and Wardle (1998) have suggested the existence of
a population of cool, self-gravitating clouds in the galactic
halo to account for ``Extreme Scattering Events'' which
occur in monitoring the flux of compact radio sources.  
They propose clouds of approximately a Jovian mass with
a radius of about an AU, which would correspond to
a gravitational acceleration 
$g \sim 6\times 10^{-4}\:\cmss$ at the ``surface.''  With 
a total column density of $\sim 10^{26}\:{\rm cm}^{-2}$, 
the majority of the AGN energy flux would be absorbed well 
before reaching (and disrupting) the center of the cloud.  
The proposed particle number density of $\sim 10^{12}\:{\rm cm}^{-3}$
corresponds well to that inferred by Ferland \etal (1996) and
Baldwin \etal (1996).  Thus if these self-gravitating, 
sub-stellar clouds exist in host galaxies of AGN, they 
seem to be potential candidates for the origin of the BLR\@.
It is unclear, though, how they could exist with a high metal content, 
as required by BLR observations, without rapidly 
cooling and collapsing (in much less than a Hubble time)
prior to exposure to AGN irradiation. 

	Although we conclude that X-ray induced \lya\ pressure 
would not be significant for main sequence stars near AGN, 
it could be important either for lower gravity 
stars which have evolved off the main sequence up the giant branch,
or for substellar, self-gravitating clouds.
It remains to be seen whether the resulting cometary 
structures can account for the required BLR covering factor 
and cloud population, and the AGN mass supply.

\acknowledgements

We gratefully acknowledge Bohdan Paczy\'nski for useful 
and inspiring discussions, Gary Ferland for help and advice
with CLOUDY, and R. H. Lupton for the availability of the
SM plotting package.  The research was supported in part by
NSF grant AST-9619429.

\begin{figure}
\plotone{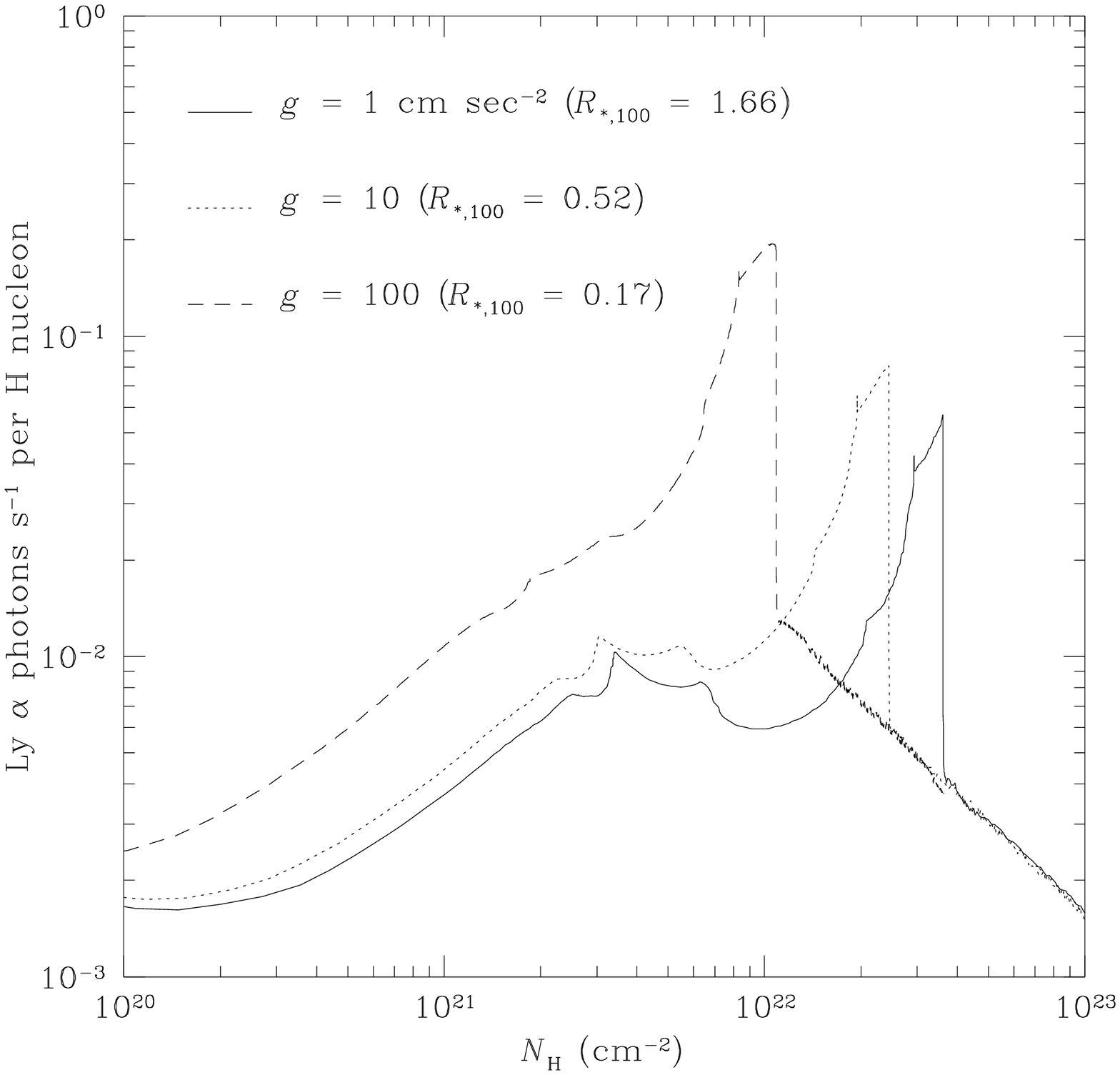}
\caption{
The \lya\ photon injection rate per H nucleon as a 
function of total hydrogen column density $N_{\rm H}$, in
the cases of $g=1,$ 10, and 100$\:\cmss$.
The incident flux is given by equation
(\ref{eq:fluxdist}) with $A/(4\pi d^2) = 2.37\times 10^{10}\:
\ergcmcms,$ corresponding to a distance 0.25$\:\hpc$ 
from an AGN of luminosity $L_{46} = 27.8\:h_{65}^{-2}$.
The heating distance $d_{h}$ for this value of $L_{46}$
is $0.26\:\hpc$.  
The atmosphere is assumed to have $n_{\rm H}=10^{10}\:{\rm cm}^{-3}$ 
at $N_{\rm H} = 0.$  
The stellar radius $R_{*,100}$ 
in units of $100R_{\sun}$ is given for each value of 
$g$, assuming a stellar mass of $M_{\sun}$.
}
\end{figure}

\begin{figure}
\plotone{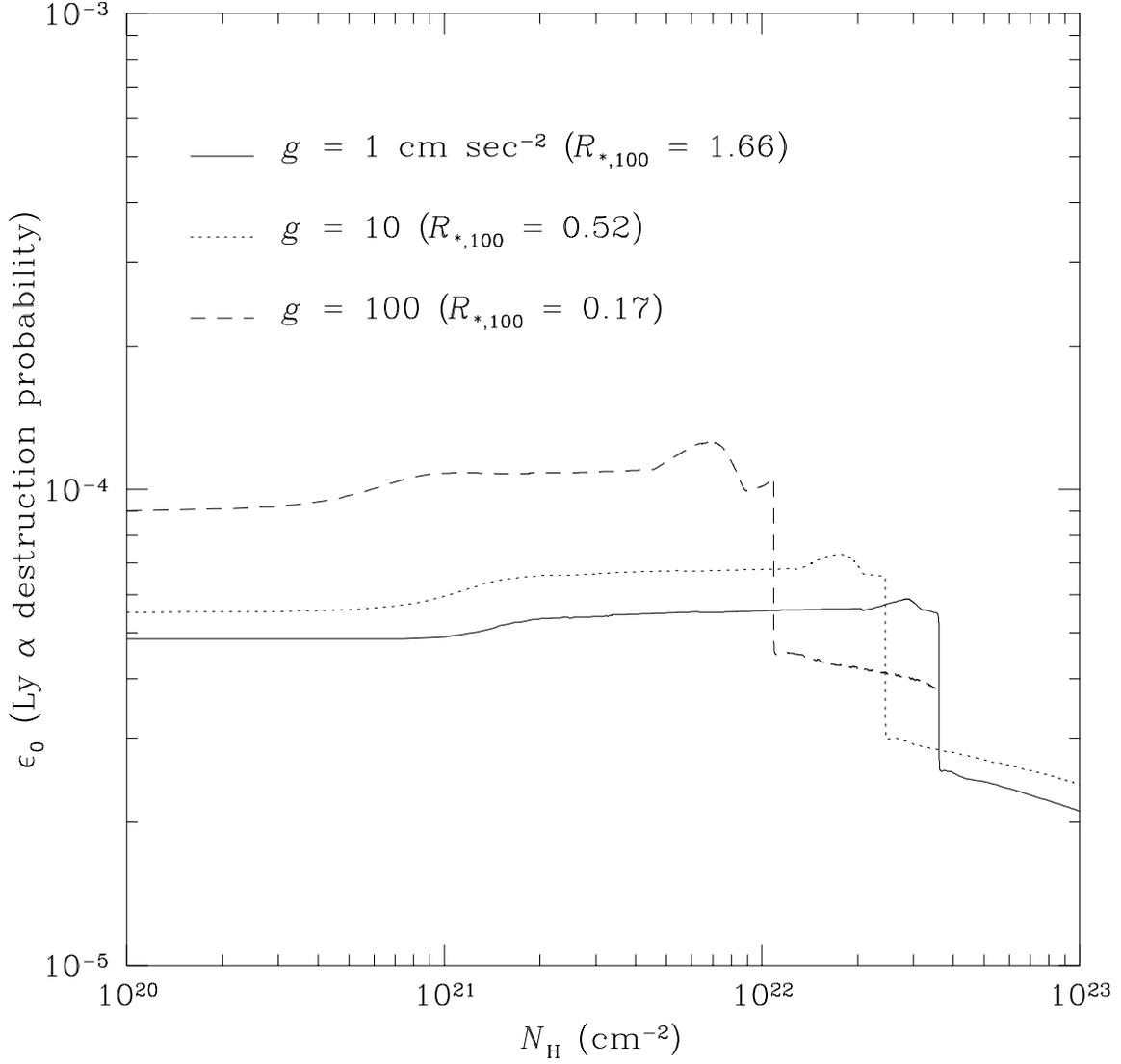}
\caption{
The \lya\ photon destruction probability per scattering $\epsilon_0$
as a function of total hydrogen column density $N_{\rm H},$ 
calculated from equation (\ref{eq:epsdef}), 
in the cases of $g=1,$ 10, and 100 $\:\cmss$.
The incident flux is given by equation
(\ref{eq:fluxdist}) with $A/(4\pi d^2) = 2.37\times 10^{10}\:
\ergcmcms.$
}
\end{figure}

\begin{figure}
\plotone{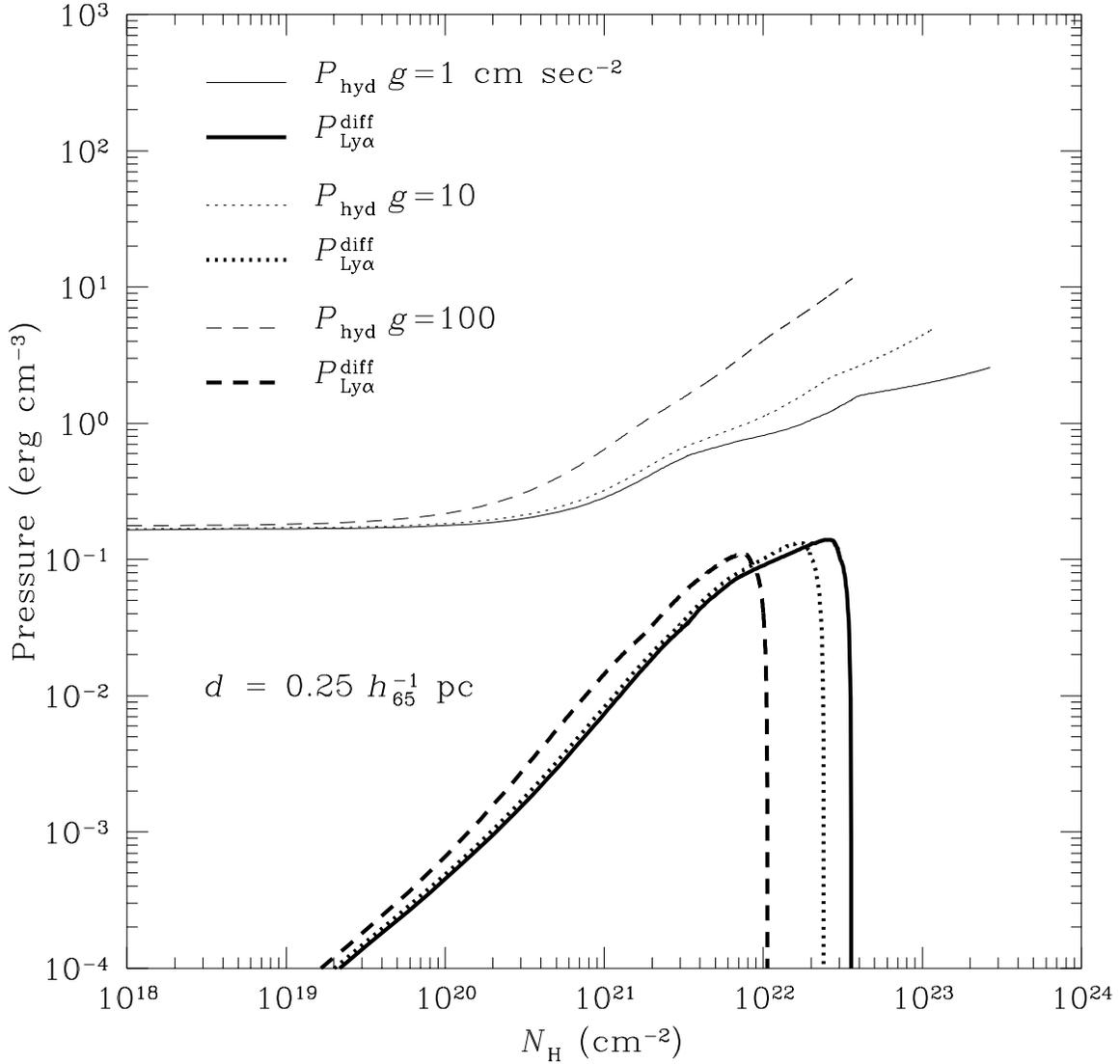}
\caption{
The differential radiation pressure $P_{\rm Ly\alpha}^{\rm diff}$ 
(thick lines) due to trapped \lya\ photons
is plotted as a function of total H column density $N_{\rm H},$
for $g = 1,\ 10,\ {\rm and}\ 100\:\cmss$.
The incident flux is given by equation (\ref{eq:fluxdist})
with $A/(4\pi d^2) = 2.37\times 10^{10}\:
\ergcmcms.$  The thin lines are the  
pressures required for hydrostatic support $P_{\rm hyd}$ 
at the same gravities [see equation (\ref{eq:p_hyddef})], 
where we have assumed a density 
$n_{\rm H} = 10^{10}\ {\rm cm^{-3}}$ at $N_{\rm H} = 0$.  
}
\end{figure}

\begin{figure}
\plotone{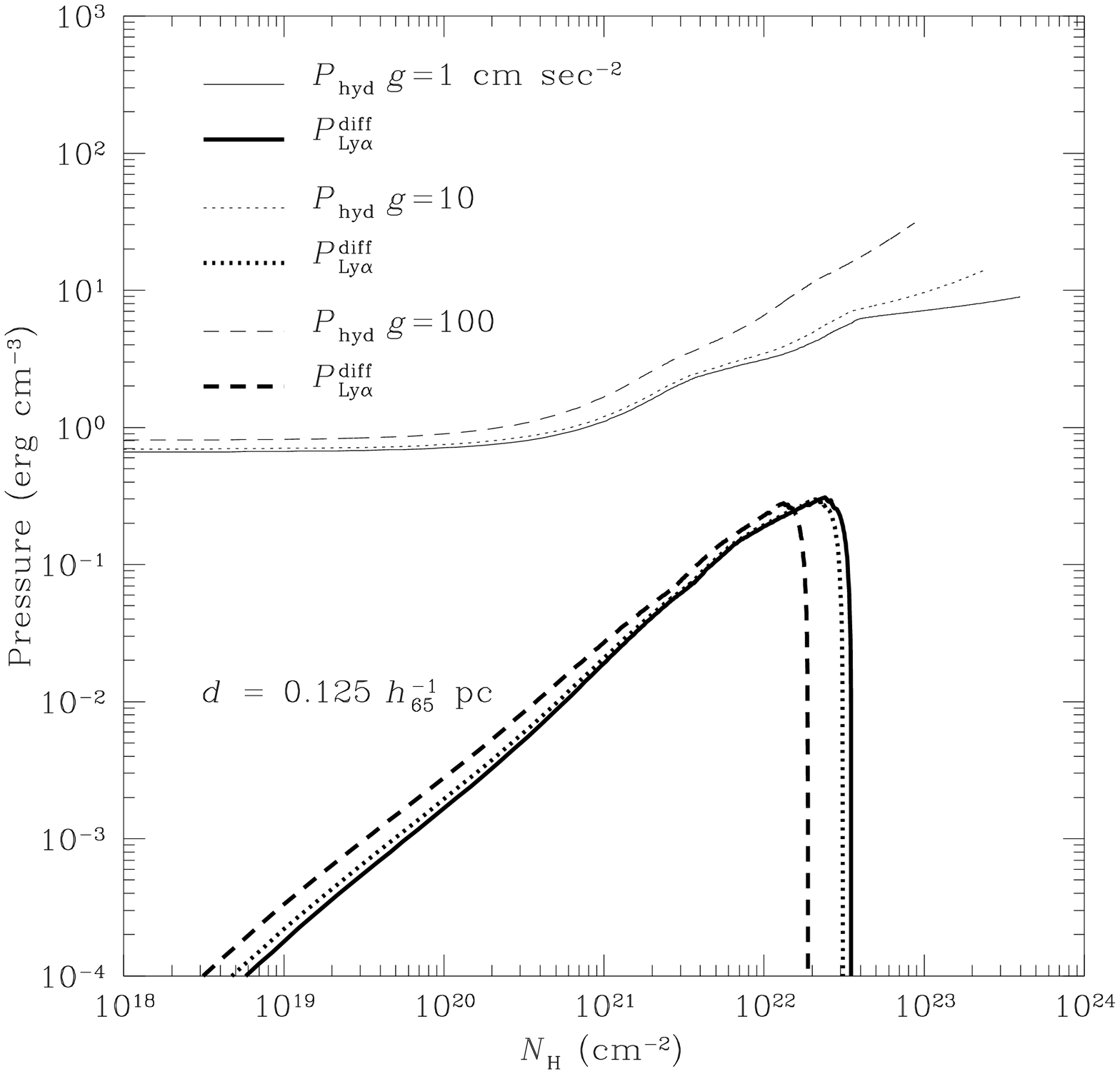}
\caption{
Same as Figure 3, but for $A/(4\pi d^2) = 9.47\times 10^{10}\:
\ergcmcms,$ corresponding to 
a distance $d=0.125\:\hpc$ 
from an AGN with $L_{46} = 27.8\: h_{65}^{-2}.$
}
\end{figure}

\begin{figure}
\plotone{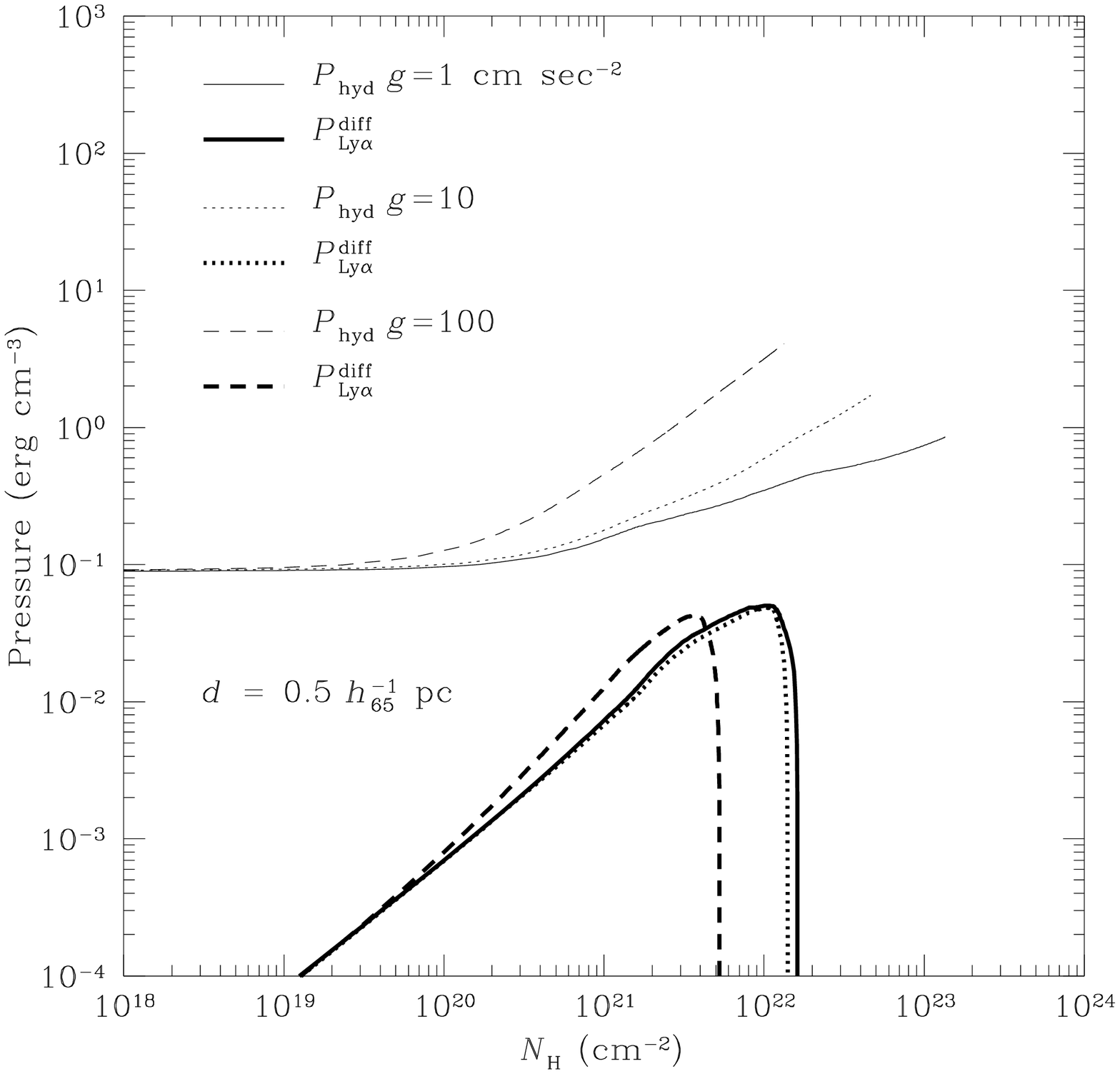}
\caption{
Same as Figure 3, but for $A/(4\pi d^2) = 5.92\times 10^{9}\:
\ergcmcms,$ corresponding to 
a distance $d=0.5\:\hpc$ from 
an AGN with $L_{46} = 27.8\: h_{65}^{-2}.$
}
\end{figure}

\begin{figure}
\plotone{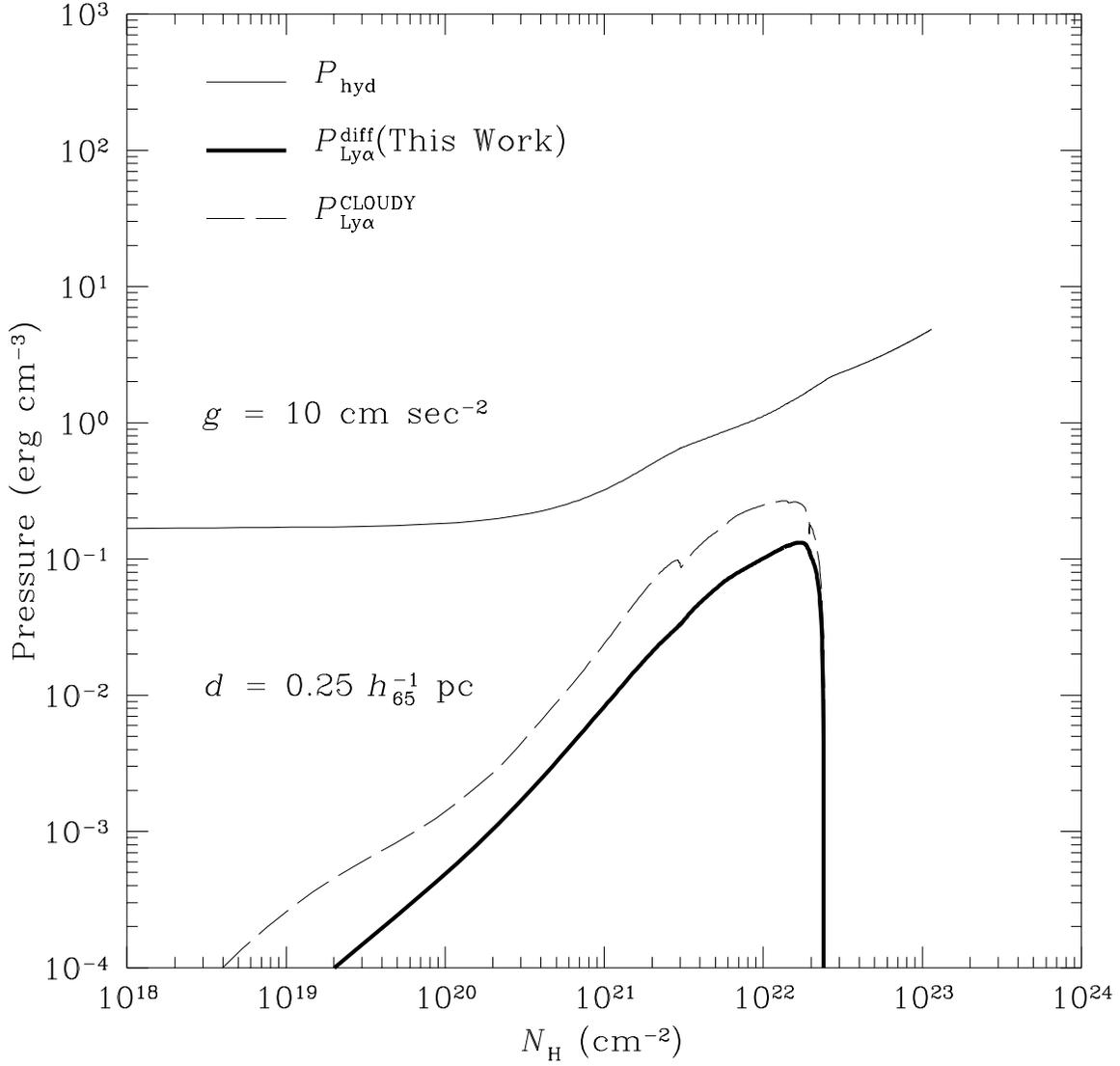}
\caption{
Comparison of the \lya\ pressure derived from our calculation
with that calculated by CLOUDY in the escape probability
approximation, for $g=10\:\cmss$, and
$A/(4\pi d^2) = 2.37\times 10^{10}\:
\ergcmcms$ as in Figures~1--3.
}
\end{figure}

\begin{figure}
\plotone{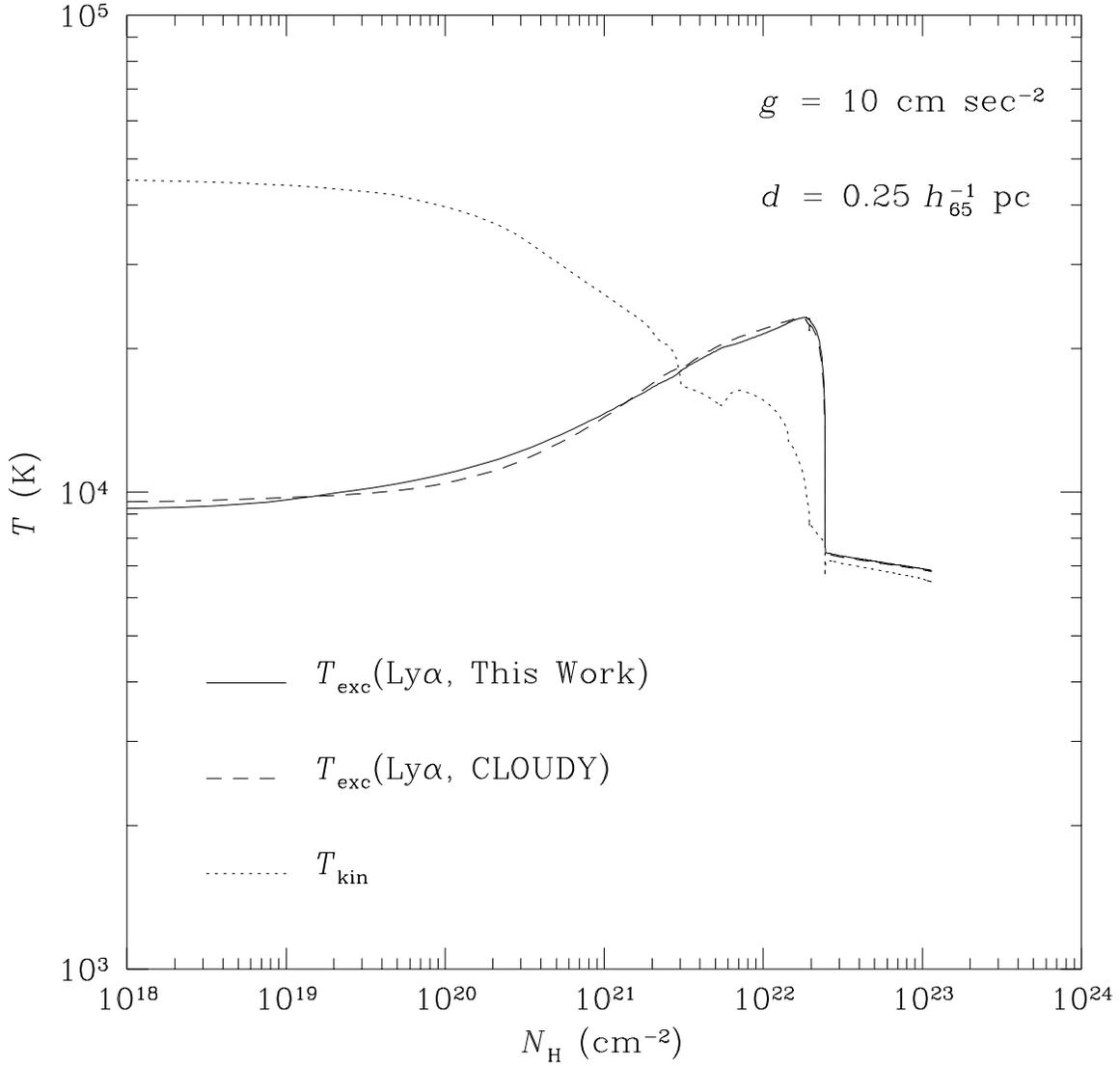}
\caption{
Comparison of the excitation temperature $T_{\rm exc}$
in \lya\ derived from our calculation
with that calculated by CLOUDY in the escape probability
approximation, for $g=10\:\cmss$, and 
$A/(4\pi d^2) = 2.37\times 10^{10}\:
\ergcmcms$ as in Figures~1--3 and 6.
For reference, the kinetic
temperature $T_{\rm kin}$ is also given.
}
\end{figure}

\end{document}